\documentclass[12pt]{spieman}  
\usepackage{amsmath,amsfonts,amssymb}
\usepackage{graphicx}
\usepackage{setspace}
\usepackage{tocloft}
\usepackage{lineno}
\usepackage{color}

\title{Maunakea Spectroscopic Explorer exposure time calculator for end-to-end simulator: to optimizing spectrograph design and observing simulation}

\author[a]{Tae-Geun Ji}
\author[b]{Jennifer Sobeck}
\author[a]{Changgon Kim}
\author[a]{Hojae Ahn}
\author[d,e]{Mingyeong Yang}
\author[c]{Taeeun Kim}
\author[d,e]{Sungwook E. Hong}
\author[b,f]{Kei Szeto}
\author[g]{Jennifer L. Marshall}
\author[h]{Christian Surace}
\author[a,*]{Soojong Pak}
\affil[a]{Kyung Hee University, School of Space Research, 1732 Deogyeong-daero, Giheung-gu, Yongin-si Gyunggi-do, Republic of Korea, 17104}
\affil[b]{CFHT Corporation, 65-1238 Mamalahoa Hwy, Kamuela, HI, USA, 96743}
\affil[c]{Kyung Hee University, Department of Astronomy and Space Science, 1732 Deogyeong-daero, Giheung-gu, Yongin-si Gyunggi-do, Republic of Korea, 17104}
\affil[d]{Korea Astronomy and Space Science Institute, Cosmology Group, 776 Daedeok-daero, Yuseong-gu, Daejeon, Republic of Korea, 34055}
\affil[e]{University of Science \& Technology, Astronomy Campus, 217 Gajeong-ro, Yuseong-gu, Daejeon, Republic of Korea, 34113}
\affil[f]{National Research Council Canada, Herzberg Astronomy and Astrophysics, 5071 West Saanich Road, Victoria, BC, Canada, V9E 2E7}
\affil[g]{Texas A\&M University, Mitchell Institute for Fundamental Physics and Astronomy and Department of Physics and Astronomy, 4242 TAMU 576 University Dr Mitchell Institute Building, College Station, TX, USA, 77843}
\affil[h]{Aix-Marseille Université, Laboratoire d’Astrophysique de Marseille, Jardin du Pharo, 58 Boulevard Charles Livon, Marseille, France, 13007}

\cftpagenumbersoff{figure}
\cftpagenumbersoff{table} 
\begin{document}
\maketitle

\begin{abstract}
The Maunakea Spectroscopic Explorer (MSE) project will provide multi-object spectroscopy in the optical and near-infrared bands using an 11.25-m aperture telescope, repurposing the original Canada-France-Hawaii Telescope (CFHT) site. MSE will observe 4,332 objects per single exposure with a field of view of 1.5 square degrees, utilizing two spectrographs with low-moderate (R$\sim$3,000, 6,000) and high (R$\approx$30,000) spectral resolution. In general, an exposure time calculator (ETC) is used to estimate the performance of an observing system by calculating a signal-to-noise ratio (S/N) and exposure time. We present the design of the MSE exposure time calculator (ETC), which has four calculation modes (S/N, exposure time, S/N trend with wavelength, and S/N trend with magnitude) and incorporates the MSE system requirements as specified in the Conceptual Design. The MSE ETC currently allows for user-defined inputs of target AB magnitude, water vapor, airmass, and sky brightness AB magnitude (additional user inputs can be provided depending on computational mode). The ETC is built using Python 3.7 and features a graphical user interface that allows for cross-platform use. The development process of the ETC software follows an Agile methodology and utilizes the Unified Modeling Language (UML) diagrams to visualize the software architecture. We also describe the testing and verification of the MSE ETC.
\end{abstract}


{\noindent \footnotesize\textbf{*}Soojong Pak,  \linkable{soojong@khu.ac.kr} }

\begin{spacing}{2}   

\section{Introduction}
\label{sec:intro}  
The Maunakea Spectroscopic Explorer (MSE) will be a dedicated survey facility, simultaneously executing a wide variety of scientific programs.  Consequently, MSE will be able to perform transformative science and address a myriad of key scientific questions. The MSE Conceptual Design (CoD) involves multi-object spectroscopy (MOS) for 4,332 objects over a 1.52 square degree field of view, utilizing a primary mirror with an overall diameter of 11.25 meters. MSE will employ two spectroscopic instruments of low/moderate resolution (LMR; R$\sim$3,000, 6,000) and high resolution (HR; R$\approx$30,000) with multiplexing allocations of 3,249 and 1,083 fibers respectively\cite{Szeto2020a}.

An exposure time calculator (ETC) generates quick estimations of the signal-to-noise ratio (S/N) and exposure time required for observations of astronomical objects with a particular telescope and instrument setup (and can be used for observing time requests). Further, an ETC can be used in the analysis of instrument performance in the design phase as well as be employed in the evaluation of the scientific programs in the development of targeting strategy and survey design\cite{Pontoppidan2016,Nielsen2016,Glover2022}. For instrument evaluation and verification, an ETC provides initial estimates of system performance while an End-to-End (E2E) simulation generates detailed simulations of system performance and astronomical observations\cite{Nord2016,Genoni2020,Briesemeister2020}. Considerations in the design of the ETC include: (1) determination of the target types as based on predetermined scientific objectives, (2) incorporation of the specified instrument parameters (such as system throughput) and (3) adoption of an appropriate sky model for sky background estimation.

MOS observations enable the swift execution and the accumulation of large samples for astronomical survey programs (as compared, for example, to single-slit observations)\cite{MacKenty2003,Ellis2017}. It is highly challenging for an ETC to generate estimates for all MOS targets within a single field.  Currently, the MSE ETC simulates a single target of a given brightness and exposure time. Multiple runs of the ETC are therefore necessary to simulate the range of targets in a particular field (these run data can be used in turn to inform targeting strategy and survey design).

For ETC software of ground-based observatories, consideration of atmospheric extinction and determination of the sky background (which varies as a function of wavelength) must also occur\cite{Burke2010}. In this regard, the MSE ETC is informed by previously-developed ETC code such as the ETC for the Immersion GRating INfrared Spectrograph (IGRINS)\cite{Huynh2015}, which employs the OH infrared emission line data from Ref.~\citenum{Rousselot2000}. The ETC also relies upon the ESO SkyCalc tool for sky-atmosphere modeling, which is also used by (for example) the ETC of Son Of X-Shooter (SOXS)\cite{Genoni2022}.

In this paper, we describe the design of the MSE ETC, including algorithms, architecture, and verification. In Section \ref{sec:mse}, we briefly present an overview of the MSE instrument as well as how the MSE ETC fits into the 
collection of high-level software modules that will be relied upon to facilitate science operations. The overall algorithm of the MSE ETC is discussed in Section \ref{sec:algorithm}. Section \ref{sec:parameters} presents the instrument and throughput parameters of the MSE system. Section \ref{sec:methods} describes the simulation methods used in each computation mode. Section \ref{sec:software} shows the development process, architecture, and graphical user interface (GUI) of the MSE ETC. Section \ref{sec:results} describes the simulation results, and finally, conclusions are discussed in Section \ref{sec:conclusion}.

\section{Overview of MSE}
\label{sec:mse}
\subsection{The MSE Instrument}
The MSE CoD calls for a primary mirror consisting of sixty hexagonal segments with an overall diameter of 11.25 meters.  MSE will replace the existing 3.6-m CFHT facility (with no footprint extension). The MSE low/moderate resolution (LMR) instrument will have a spectral resolution range from R$\sim$3,000 to R$\sim$6,000 with a wavelength range from 360 nm to 1800 nm. The MSE high resolution (HR) instrument will have an approximate resolution of R$\approx$30,000 and wavelength coverage of about 360 nm to 900 nm\cite{Zhang2020a}. Table~\ref{tab:table_resolutions} shows the detailed spectral coverage and resolution of the LR, MR, and HR spectroscopic settings for the CoD design \cite{Hill2018}. The spectral resolutions are determined at the central wavelength of each band---482 nm, 626 nm, and 767 nm correspond to the effective wavelength of the \textit{g'}, \textit{r'}, and \textit{i'} bands of the Sloan Digital Sky Survey (SDSS) filter system\cite{Fukugita1996}. In addition, the 1235 nm central wavelength of the NIR band for the LR instrument corresponds to the effective wavelength of the $J$ band of the Two Micron All Sky Survey (2MASS)\cite{Cohen2003}.

\begin{table} [t]
\caption{Spectral resolutions of the LR, MR, and HR spectroscopy in the full wavelength range.}
\label{tab:table_resolutions}
\vspace{-0.4cm}
\begin{center}
\resizebox{0.82\textwidth}{!}
{\tiny
\begin{tabular}{ccccc}
\noalign{\smallskip}\noalign{\smallskip}
\hline\hline
\multicolumn{5}{c}{Low Resolution (LR) spectroscopy} \\ 
\hline
\multicolumn{1}{l|}{Band channel} 
& Blue & Green & Red & NIR\\
\multicolumn{1}{l|}{Spectral resolution} 
& 2,550 & 3,650 & 3,600 & 3,600\\
\multicolumn{1}{l|}{Wavelength range (nm)}
& 360--560 & 540--740 & 715--985 & 960--1800 \\
\multicolumn{1}{l|}{Center wavelength (nm)}
& 482 & 626 & 900 & 1235\\
\hline\hline
\multicolumn{5}{c}{Moderate Resolution (MR) spectroscopy} \\ 
\hline
\multicolumn{1}{l|}{Band channel} 
& Blue & Green & Red & \\
\multicolumn{1}{l|}{Spectral resolution} 
& 4,400 & 6,200 & 6,100 & \\
\multicolumn{1}{l|}{Wavelength range}
& 391--510 & 576--700 & 737--900 & \\
\multicolumn{1}{l|}{Center wavelength}
& 482 & 626 & 767 & \\
\hline\hline
\multicolumn{5}{c}{High Resolution (HR) spectroscopy} \\ 
\hline
\multicolumn{1}{l|}{Band channel} 
& Blue & Green & Red & \\
\multicolumn{1}{l|}{Spectral resolution} 
& 40,000 & 40,000 & 20,000 & \\
\multicolumn{1}{l|}{Wavelength range}
& 360--460 & 440--620 & 600--900 & \\
\multicolumn{1}{l|}{Center wavelength}
& 400 & 482 & 767 & \\
\hline\hline
\end{tabular}
}
\end{center}
\end{table}

\subsection{MSE Survey Program Execution}

The execution of MSE survey programs will be facilitated by an Operations Model and the supporting Program Execution System Architecture (PESA). PESA will be the end-to-end high-level software suite that involves various aspects of MSE science program execution, from proposal submission to the distribution of the science data\cite{Szeto2020b}. The MSE ETC will be contained within the Survey Preparation and Definition (SPD) group in the pre-observation phase of the PESA platform\cite{Szeto2020b}.

\section{Main ETC Algorithm}
\label{sec:algorithm}

\begin{figure} [t]
\includegraphics[width=\textwidth]{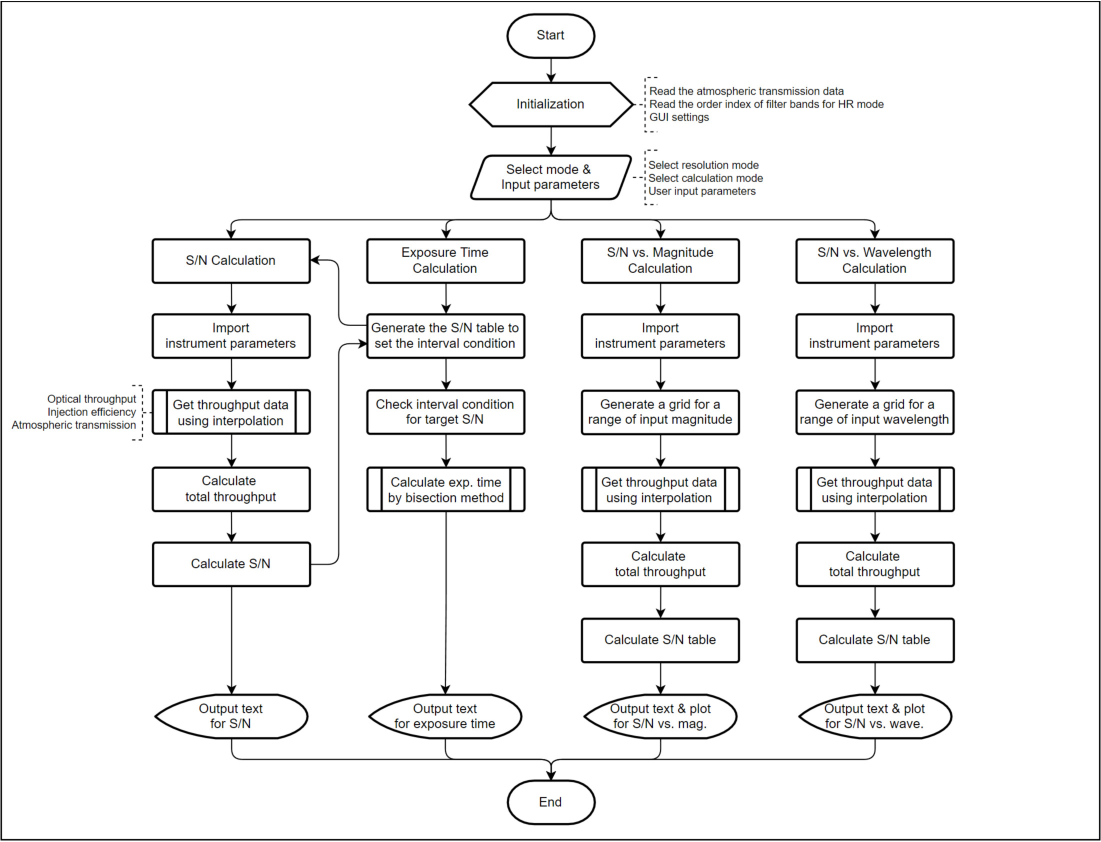}
\caption
{ \label{fig:overall}
Processing algorithm of the MSE ETC. There are four calculation methods: S/N, exposure time, S/N as a function of magnitude, and S/N as a function of wavelength. Each method runs individually and has its own output channel.}
\end{figure}
We have designed the algorithm of the MSE ETC and have written the associated code in Python (with a few external package dependencies). The ETC runs locally, and users launch the GUI through the command-line interface (CLI). Figure \ref{fig:overall} displays a flowchart of the code and accordingly provides a brief overview of the functionalities and data flow. The general code flow of the MSE ETC is as follows: initialization of the pre-defined parameters; selection of the desired spectral resolution setting and specification of certain input parameters by the user (target AB magnitude, sky background AB magnitude, precipitable water vapor, and airmass; default values are set in the code), user selection of the calculation mode and execution of the calculation sequence (once all parameters are input), and generation of the associated calculation output. Note that the choice of spectral resolution sets a variety instrumental parameters, such as the optical throughput and accessible wavelength bands. There are four calculation modes of the MSE ETC: S/N calculation, exposure time calculation, S/N vs. Magnitude, and S/N vs. Wavelength. Each calculation mode executes input and output operations independently. Depending on calculation mode, additional user inputs can include: exposure time, S/N, and number of exposures.

At the initialization phase, the MSE ETC reads in inputs related to GUI settings and the atmospheric transmission data (which is saved as a list; this data is employed in the estimate of total throughput). Note that for the HR spectrograph, only one order per spectrograph channel is available for data acquisition at once.  Accordingly, the user must select among a pre-defined set of HR spectrograph bandpasses for each of the three channels (these bandpasses are labeled as indexes). The selection of HR bandpasses settings occurs in the combo box of the GUI. After initialization, reading of the desired calculation mode and other various user inputs occurs. The ETC then can execute the chosen calculation.

We now briefly describe each of the calculation modes below.  The S/N calculation mode generates single S/N values for the central wavelength of each band associated with the selected spectral resolution setting. This calculation involves importing the instrumental parameters, obtaining the total throughput data, computing the S/N, and generating the output results as text. The exposure time calculation mode provides the total exposure time for the central wavelength of each band associated with the chosen spectral resolution setting. This mode calls the S/N calculation for generating the table of S/N values and then estimates the exposure time using the bisection method. The S/N vs. Magnitude mode produces a plot of the S/N (for a specified target AB magnitude) as a function of the magnitude range for each band of the chosen spectral resolution (the generated values are also provided to users). Similarly, the S/N vs. Wavelength mode provides a plot and generated data for the correlation between the S/N and the full wavelength range for each band at the chosen spectral resolution.

\section{Parameters}
\label{sec:parameters}
Before explaining the simulation methods, we define the parameters required for the S/N calculation, such as the instrument specifications, the optical throughput for the MSE system, the atmospheric transmission, and the OH emission.

\subsection{Instrumental Parameters}
The instrumental parameters refers to the as-designed specifications of the MSE instruments (as set in the CoD). Table~\ref{tab:table_telescope} shows the telescope, fiber, and detector parameters. In the telescope, $D_{\text{tel\_eff}}$ and $A_{\text{tel}}$ represent the effective aperture and light-collecting area, respectively. The $\Omega_{\text{fiber}}$ is the solid angle viewed on the sky with a fiber slit size of 1 arcsecond. The $n_{\text{res}}$ represents the spectral and spatial resolution elements of the detector. We also consider the dark current noise ($N_{\text{dark}}$) and read noise ($N_{\text{read}}$) of the detector in the calculation. The thermal background noise ($N_{\text{thermal}}$) is also critical to evaluate realistic S/N calculation in NIR regime. The temperature and total emissivity of the instrument components determine the number of thermal electrons generated on the detector\cite{Huynh2015}. Note that the current allocation only uses the total thermal emission on the NIR detector in LR spectrograph.

\begin{table} [!ht]
\caption{Parameters for telescope, fiber and detector.}
\label{tab:table_telescope}
\vspace{-0.2cm}
\begin{center}
\resizebox{\textwidth}{!}
{
\begin{tabular}{cccc}
\noalign{\smallskip}\noalign{\smallskip}
\hline\hline
\multicolumn{1}{l}{Parameter} 
& \multicolumn{1}{l}{Description}
& \multicolumn{1}{l}{Value}
& \multicolumn{1}{l}{Unit} \\
\hline
\multicolumn{1}{l}{$D_{\text{tel\_eff}}$}
& \multicolumn{1}{l}{Telescope effective aperture}
& \multicolumn{1}{l}{10.14}
& \multicolumn{1}{l}{m} \\
\multicolumn{1}{l}{$A_{\text{tel}}$}
& \multicolumn{1}{l}{Telescope light-collecting area}
& \multicolumn{1}{l}{80.75}
& \multicolumn{1}{l}{$\text{m}^2$} \\
\multicolumn{1}{l}{$\Omega_{\text{fiber}}$}
& \multicolumn{1}{l}{Solid angle viewed on the sky with $1''$ fiber size}
& \multicolumn{1}{l}{0.785}
& \multicolumn{1}{l}{$\text{arcsec}^{2}$} \\
\multicolumn{1}{l}{$n_{\text{res}}$}
& \multicolumn{1}{l}{Spectral and spatial resolution elements of detector}
& \multicolumn{1}{l}{4$\times$4 (16)}
& \multicolumn{1}{l}{pixel} \\
\multicolumn{1}{l}{$N_{\text{dark}}$}
& \multicolumn{1}{l}{Detector dark current noise}
& \multicolumn{1}{l}{0.02}
& \multicolumn{1}{l}{electron $\text{s}^{-1}$} \\
\multicolumn{1}{l}{$N_{\text{read}}$}
& \multicolumn{1}{l}{Detector read noise}
& \multicolumn{1}{l}{5 (R, G, B), 8 (NIR)}
& \multicolumn{1}{l}{electron} \\
\multicolumn{1}{l}{$N_{\text{thermal}}$}
& \multicolumn{1}{l}{Total thermal emission in NIR band}
& \multicolumn{1}{l}{9}
& \multicolumn{1}{l}{electron $\text{s}^{-1}$} \\
\hline\hline
\end{tabular}
}
\end{center}
\end{table}

\subsection{Throughput Parameters}
\label{subsec:throughput}
Multiple factors influence the throughput determination, which include light transmission through the telescope structure, the spectrograph, and the atmosphere. We define the total throughput ($\tau_{\text{total}}$) as the product of the optical throughput ($\tau_{\text{opt}}$), the injection efficiency ($\tau_{\text{ie}}$), and atmospheric transmission ($\tau_{\text{atmo}}$) in Equation (\ref{tau_total}). The estimates for certain aspects of the system throughput (over the full MSE wavelength range) as well as the injection efficiency are taken from the MSE Sensitivity Budget Allocation document\cite{Flagey2018a}.

\begin{equation}\label{tau_total}
\tau_{\text{total}}=\tau_{\text{opt}}\times\tau_{\text{ie}}\times\tau_{\text{atmo}}
\end{equation}

\subsubsection{Optical Throughput and Injection Efficiency}

The optical throughput is estimated by considering the light transmission that occurs through each MSE subsystem. Table~\ref{tab:factors_opt} lists all of the considered MSE subsystems.
\begin{table}[b]
\begin{center}
\caption{Factors in the MSE optical throughput.}
\label{tab:factors_opt}
\resizebox{\textwidth}{!}
{
\begin{tabular}{ccc}
\noalign{\smallskip}\noalign{\smallskip}
\hline\hline
\multicolumn{1}{l}{Subsystem} 
& \multicolumn{1}{l}{Description}
& \multicolumn{1}{l}{Value} \\
\hline
\multicolumn{1}{l}{ENCL}
& \multicolumn{1}{l}{Telescope enclosure}
& \multicolumn{1}{l}{1.000} \\
\multicolumn{1}{l}{M1, ZeCoat}
& \multicolumn{1}{l}{Primary mirror reflectivity with ZeCoat protected silver coating}
& \multicolumn{1}{l}{0.94--0.98} \\
\multicolumn{1}{l}{MSTR}
& \multicolumn{1}{l}{Telescope mount structure}
& \multicolumn{1}{l}{0.96} \\
\multicolumn{1}{l}{PFHS}
& \multicolumn{1}{l}{Prime focus haxapod system}
& \multicolumn{1}{l}{0.99} \\
\multicolumn{1}{l}{WFC/ADC}
& \multicolumn{1}{l}{Wide field corrector and atmospheric dispersion corrector}
& \multicolumn{1}{l}{0.52--0.85} \\
\multicolumn{1}{l}{PosS}
& \multicolumn{1}{l}{Fibre positioner system}
& \multicolumn{1}{l}{0.97} \\
\multicolumn{1}{l}{FiTS}
& \multicolumn{1}{l}{Fibre transmission system}
& \multicolumn{1}{l}{0.57--0.87 (LR/MR)} \\
\multicolumn{1}{l}{}
& \multicolumn{1}{l}{}
& \multicolumn{1}{l}{0.43--0.88 (HR)} \\
\multicolumn{1}{l}{SIP}
& \multicolumn{1}{l}{Spectrographs, either low/moderate or high resolution}
& \multicolumn{1}{l}{0.24--0.54 (LR)} \\
\multicolumn{1}{l}{}
& \multicolumn{1}{l}{}
& \multicolumn{1}{l}{0.48--0.63 (MR without grating)} \\
\multicolumn{1}{l}{}
& \multicolumn{1}{l}{}
& \multicolumn{1}{l}{0.40--0.48 (MR with grating)} \\
\multicolumn{1}{l}{}
& \multicolumn{1}{l}{}
& \multicolumn{1}{l}{0.27--0.44 (HR without grating)} \\
\multicolumn{1}{l}{}
& \multicolumn{1}{l}{}
& \multicolumn{1}{l}{0.35 (HR with grating)} \\
\hline\hline
\end{tabular}
}
\end{center}
\end{table}
   \begin{figure} [t]
   \begin{center}
   \begin{tabular}{c} 
   \includegraphics[width=\textwidth]{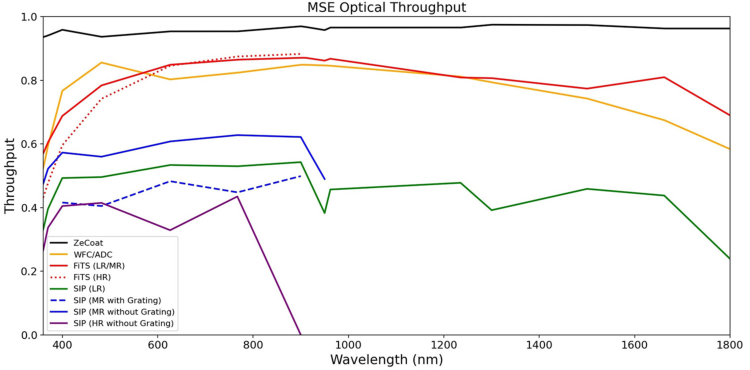}
   \end{tabular}
   \end{center}
   \caption[MSE optical throughput for each component as a function of wavelength in LR, MR, and HR modes.] 
   { \label{fig:optical_tau} MSE optical throughput for each component as a function of wavelength in LR, MR, and HR spectroscopy.}
   \end{figure}
In the design of MSE, the Telescope Enclosure (ENCL), the Telescope Mount Structure (MSTR), the Prime Focus Hexapod System (PFHS), and the Fiber Positioner System (PosS) have constant values over the full wavelength ranges of the instruments, while the others (Primary Mirror Reflectivity [M1], Wide Field Corrector [WFC], and Spectrographs [SIP]) vary with wavelength. We apply linear interpolation via scipy.interpolate.interp1d in Python to obtain the throughput values at each wavelength (see Figure \ref{fig:optical_tau}). Finally, the optical throughput is expressed by the multiplication of all of the aforementioned factors.

The Injection Efficiency (IE) is the fraction of light from a point source that reaches a fiber at the focal plane. The IE depends on the aperture size of the fiber and the image quality (which is represented by the point spread function of the telescope and the atmosphere). The IE also depends on the scale change on the focal plane, and this factor is included in the PSF evaluation\cite{Flagey2018b}. Note that vignetting due to the wide field corrector is not taken into consideration. Currently, we consider only a single point target and use the IE values at the optimal focus on the center of the focal plane. We employed the IE data from the MSE Sensitivity Budget Allocation document and then, estimated the IE at each wavelength using the linear interpolation from scipy.interpolate.interp1d in Python. The values of the IE with each spectroscopy are shown in Table~\ref{tab:factors_ie}.

\begin{table}[!ht]
\begin{center}
\caption{Injection efficiency of the LR, MR, and HR spectroscopy.}
\label{tab:factors_ie}
\resizebox{5.0cm}{!}
{
\begin{tabular}{cc}
\noalign{\smallskip}\noalign{\smallskip}
\hline\hline
\multicolumn{1}{l}{Spectral Resolution} 
& \multicolumn{1}{l}{Value} \\
\hline
\multicolumn{1}{l}{LR}
& \multicolumn{1}{l}{0.52--0.69} \\
\multicolumn{1}{l}{MR}
& \multicolumn{1}{l}{0.55--0.66} \\
\multicolumn{1}{l}{HR}
& \multicolumn{1}{l}{0.37--0.47} \\
\hline\hline
\end{tabular}
}
\end{center}
\end{table}

\subsubsection{Atmospheric Transmission}
\begin{figure} [b]
   \begin{center}
   \begin{tabular}{c} 
   \includegraphics[width=\textwidth]{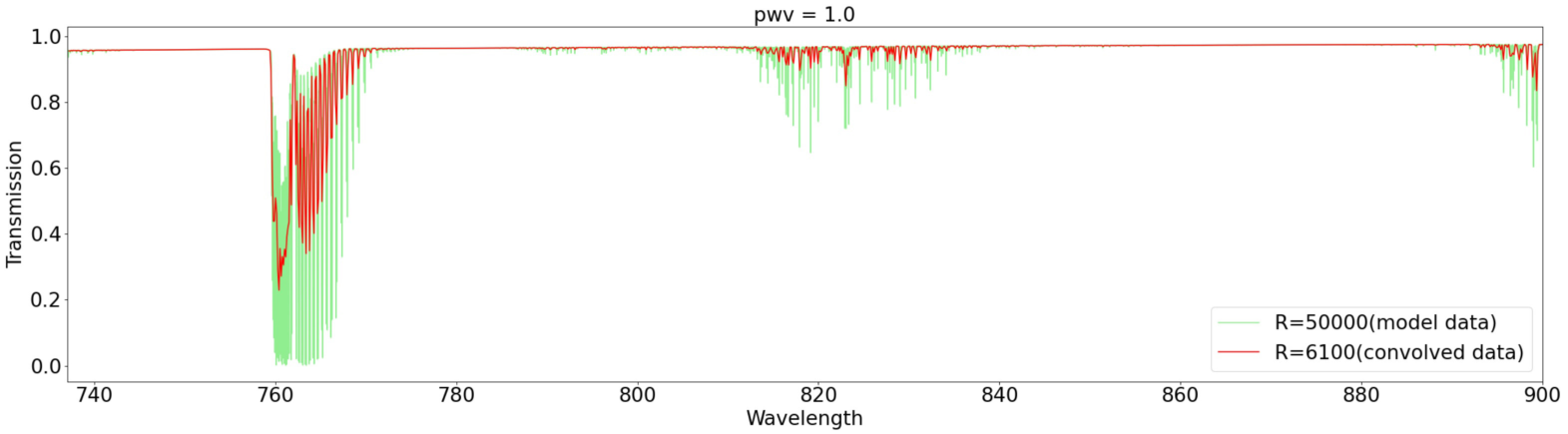}
   \end{tabular}
   \end{center}
   \caption[Sample data convolution in the red band (737$\sim$900 nm) of the MR resolution mode for the PWV 1.0 mm. The green indicates the raw model data of R=50,000 (obtained from the ESO SkyCalc Sky Model Calculator) and the red shows the convolved data of R=6,100.] 
   { \label{fig:convolution} Sample data convolution in the red band (737$\sim$900 nm) of the MR resolution mode for the PWV 1.0 mm. The green indicates the raw model data of R=50,000 (obtained from the ESO SkyCalc Sky Model Calculator) and the red shows the convolved data of R=6,100.}
   \end{figure}

We consider the Earth's atmospheric transmission (due to telluric contamination). For the MSE ETC, we use the telluric absorption spectra of the ESO SkyCalc Sky Model Calculator (website:\linkable{https://www.eso.org/observing/etc/skycalc/}) including Cerro Paranal sky model\cite{Noll2012, Jones2013}. We extracted three sets of ESO SkyCalc model data for precipitable water vapor (PWV) levels of 1.0 mm, 2.5 mm, and 7.5 mm and a spectral resolution of R=50,000 at an airmass of 1.0. The MSE ETC reads in these downloaded model data and then reconvolves them to match the spectral resolution of the MSE instruments. For a given wavelength and PWV, the ETC will return the appropriate atmospheric transmission via three-dimensional interpolation using boxcar function by astropy.convolution.box1DKernel in Python (see Figure~\ref{fig:convolution}).

\subsubsection{OH Emission Line}
\begin{figure} [!ht]
   \begin{center}
   \begin{tabular}{c} 
   \includegraphics[height=5.3cm]{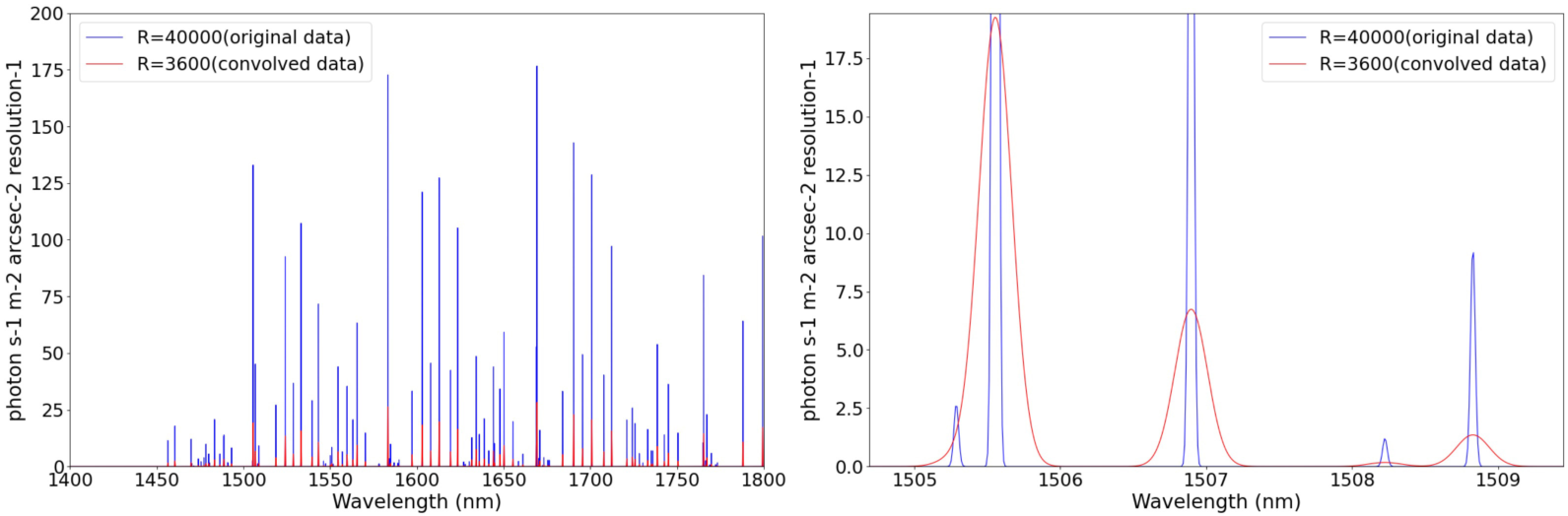}
   \end{tabular}
   \end{center}
   \caption[Plots of the OH emission lines in the MSE ETC over the wavelength range of 1,400 to 1,800 nm (left) and the zoom-in line profile (right). The blue line indicates the original data of R=40,000 (obtained from the IGRINS ETC) and the red line is the convolution spectra of R=3,600 using the Gaussian function by astropy.convolution.Gaussian1DKernel in Python.] 
   { \label{fig:OH}Plots of the OH emission lines in the MSE ETC over the wavelength range of 1,400 to 1,800 nm (left) and the zoom-in line profile (right). The blue line indicates the original data of R=40,000 (obtained from the IGRINS ETC) and the red line is the convolution spectra of R=3,600 using the Gaussian function by astropy.convolution.Gaussian1DKernel in Python.}
   \end{figure}

Currently, we calculate only the OH emission lines in the estimation of the sky background. The OH emission line data is generated by slightly modifying the OH sky line data for the H band of the Immersion GRating INfrared Spectrograph (IGRINS) ETC (i.e., reconvolving the spectra to different resolutions)\cite{Huynh2015}. The wavelengths of the OH emission lines are from Ref.~\citenum{Rousselot2000} and the calibrated intensity values are from Ref.~\citenum{Oliva1992}. The employed OH emission line data are shown in Figure~\ref{fig:OH}. A detailed description of the process for calculating the sky background with the OH lines will be provided in Section \ref{subsec:S/N}.

The line profile from the instrument which contains the dispersive element can be given by Equation (\ref{eq_inst_profile}):

\begin{equation}\label{eq_inst_profile}
I=I_{0}\left\{\frac{e^{-(\lambda-\lambda_{0})^{2}/2\sigma^{2}}a}{\sqrt{2\pi}\sigma}+\frac{(1-a)f\sigma\sqrt{2\text{log(2)}}}{\pi[(\lambda-\lambda_{0})^{2}+2f^{2}\sigma^{2}\text{log}(2)]}\right\},
\end{equation}
which is the summation of a Gaussian and a Lorentzian with parameters $a$ and $f$.\cite{Ellis2008} The contribution factor $a$ stands for the Gaussian component in the instrument profile, and the scale factor $f$ is the ratio between the FWHM of the Lorentzian component and that of the Gaussian component. Also, Ref. \citenum{Ellis2008} measured the value of $a=0.89-0.97$; The Gaussian component dominates the instrument profile, especially near the line peak, and the weak Lorentzian component arises at the wing. We selected the Gaussian function as the instrument profile approximately, and the OH sky line data is convolved with the Gaussian kernel. The instrument line profile of MSE will be measured in the future and the convolution function can be updated to reflect the measurement by introducing Equation (\ref{eq_inst_profile}) in our software.

\section{Simulation Methods}
\label{sec:methods}
\subsection{S/N Calculation} \label{subsec:S/N}
\begin{figure} [t]
   \includegraphics[width=\textwidth]{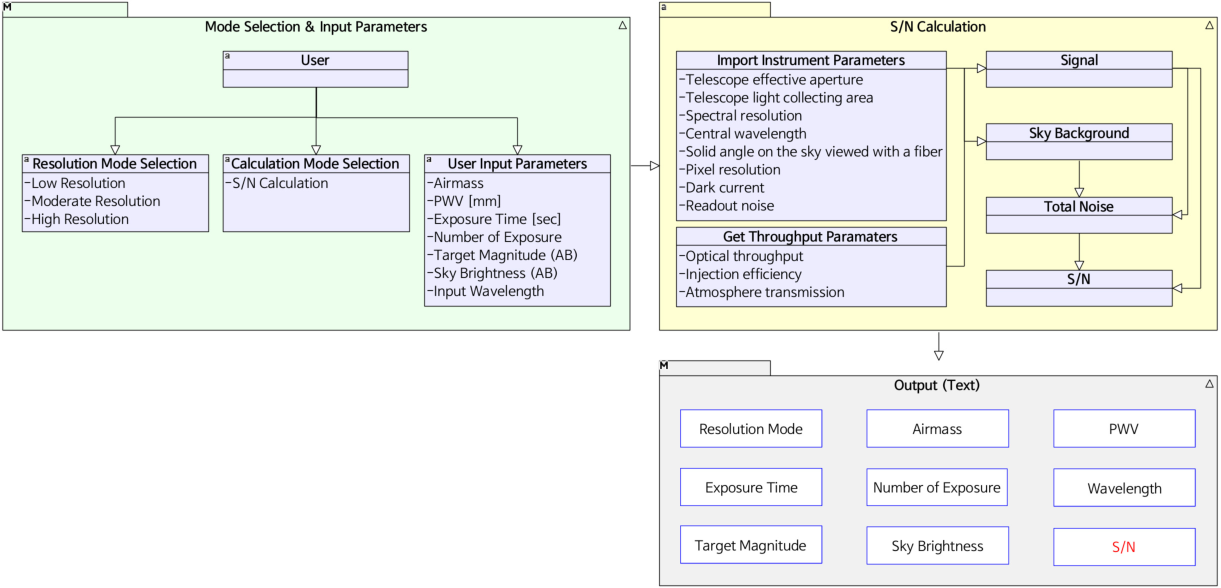}
   \caption[The flowchart shows the functions and parameters for the S/N calculation mode. The instrument, resolution, and throughput parameters are used to compute in this mode. The output terminal displays the generated S/N value together with the user input parameters.] 
   { \label{fig:fig_single} The flowchart shows the functions and parameters for S/N calculation mode. The instrument, resolution, and throughput parameters are used to compute in this mode. The output terminal displays the generated S/N value together with the user input parameters.}
   \end{figure}

The S/N calculation is designed to generate a single S/N value for a specific wavelength based on the input target magnitude. The wavelength value is fixed at the central wavelength of the band, except when the user inputs a specific wavelength. Figure~\ref{fig:fig_single} presents a detailed flowchart of this process. 

At present, the MSE ETC considers only point source targets and consequently, the formula for the continuum of a single point source target is as follows:

\begin{equation}\label{eq_signal}
S_{\text{cont}}=\frac{t_{\text{exp}}n_{\text{exp}}A_{\text{tel}}\tau_{\text{total}}10^{(m_{\text{target}}+56.1)/-2.5}}{hR},
\end{equation}

\noindent where $t_{\text{exp}}$ is exposure time for a single exposure, $n_{\text{exp}}$ is number of exposures (note that $A_{\text{tel}}$ represents the light-collecting area of the telescope, and $\tau_{\text{total}}$ is defined as the total throughput, as described in Section \ref{sec:parameters}). $m_{\text{target}}$ is a specified target AB magnitude, $h$ is Planck constant, and $R$ is spectral resolution. For the sky background, the ETC uses the formula:

\begin{equation}\label{eq_sky}
B_{\text{sky}}=\frac{t_{\text{exp}}n_{\text{exp}}A_{\text{tel}}\tau_{\text{opt}}\tau_{\text{atmo}}\Omega_{\text{fiber}}10^{(m_{\text{sky}}+56.1)/-2.5}}{hR},
\end{equation}

\noindent where $m_{\text{sky}}$ is sky brightness AB magnitude in units of $\text{mag}/\text{arcsec}^{2}$ (note that the injection efficiency is excluded because the $B_{\text{sky}}$ does not originate from the point source). For the NIR wavelength range (accessible to the LMR spectrograph), the ETC describes the OH line emission as follows:

\begin{equation}\label{eq_OH}
B_{\text{OH}}=\frac{t_{\text{exp}}n_{\text{exp}}A_{\text{tel}}\tau_{\text{opt}}\tau_{\text{atmo}}\Omega_{\text{fiber}}I_{\text{OH}}}{hR},
\end{equation}
where $I_{\text{OH}}$ is intensity of OH emission lines in units of photon $\text{s}^{-1}\text{m}^{-2}\text{arcsec}^{-2}$. In the NIR band, e.g. H-band, the background sky brightness is mainly from the OH emission lines rather than the interline continuum, mostly the zodiacal light\cite{Ellis2008}. Thus, when we use $B_{\text{OH}}$, proper consideration of $B_{\text{sky}}$ is required not to overestimate the effect of the OH emissions. Another problem related to the OH line is internal scattering. The OH line is so bright that it could be scattered inside the instrument and increase the photon counts in overall detector area\cite{Huynh2015}. These considerations are not included in this work.

For the optical wavelength range, especially the blue band, the dominant contributor of the optical sky background is the scattered moonlight in the Earth's atmosphere\cite{Noll2012}. The moonlight scattering on Earth has two types: The first type is the Rayleigh scattering from atmospheric gases, and the second type is Mie scattering by atmospheric aerosols\cite{krisciunas1991}. The scattering function $f(\rho)$ can be expressed as

\begin{equation}\label{eq_Moon_Scattering}
f(\rho) = 10^{5.36}[1.06+\text{cos}^{2}(\rho)]+10^{6.15-\rho/40},
\end{equation}

\noindent
where the two scattering terms (Note that the first term is the function of Rayleigh scattering and the second term is the function of Mie scattering) can be defined from the empirical measurement of the observation site\cite{krisciunas1991}. In this case, $\rho$ is the scattering angle for single scattering, which means the angular separation between the Moon and the target. Ref.~\citenum{krisciunas1991} suggested the lunar sky brightness model as

\begin{equation}\label{eq_moon1}
m_{\text{moon}}=f(\rho)10^{-0.4(V_{\text{moon}} + 16.57 + kX_{\text{moon}})}[1-10^{-0.4kX_{\text{target}}}],
\end{equation}

\noindent
where $V_{\text{moon}}$ is the V magnitude of the Moon, which is also a function of the phase angle and the target separation from the moon. Also, $k$ is the extinction coefficient, $X_{\text{target}}$ is the air mass of the simulated target, and $X_{\text{moon}}$ is the air mass of the moon. Finally, we can derive the lunar sky background ($B_{\text{moon}}$) as follows:

\begin{equation}\label{eq_moon2}
B_{\text{moon}}=\frac{t_{\text{exp}}n_{\text{exp}}A_{\text{tel}}\tau_{\text{opt}}\tau_{\text{atmo}}\Omega_{\text{fiber}}10^{(m_{\text{moon}}+56.1)/-2.5}}{hR},
\end{equation}

\noindent
where $m_{\text{moon}}$ is lunar sky brightness in units of $\text{mag}/\text{arcsec}^{2}$.

The total noise ($N_{\text{cont}}$) is expressed by Equation (\ref{total_noise}) (with all quantities previously defined). Finally, we arrive at the formula (Equation (\ref{snr})) for the S/N as computed by the ETC. (Note that $B_{\text{OH}}$ and $N_{\text{thermal}}$ are only considered in NIR band of LR mode. Also, $B_{\text{moon}}$ is only considered as a model and is not currently included. We may consider the Moonlight in the next versions of ETC.)

\begin{equation}\label{total_noise}
N_{\text{cont}}=\sqrt{S_{\text{cont}}+B_{\text{sky}}+B_{\text{OH}}+B_{\text{moon}}+n_{\text{res}}n_{\text{exp}}[t_{\text{exp}}(N_{\text{dark}}+N_{\text{thermal}})+N_{\text{read}}^{2}]}
\end{equation}

\begin{equation}\label{snr}
\text{S/N}=\frac{S_{\text{cont}}}{N_{\text{cont}}}
\end{equation}

\subsection{Exposure Time Calculation}

\begin{figure} [t]
   \includegraphics[width=\textwidth]{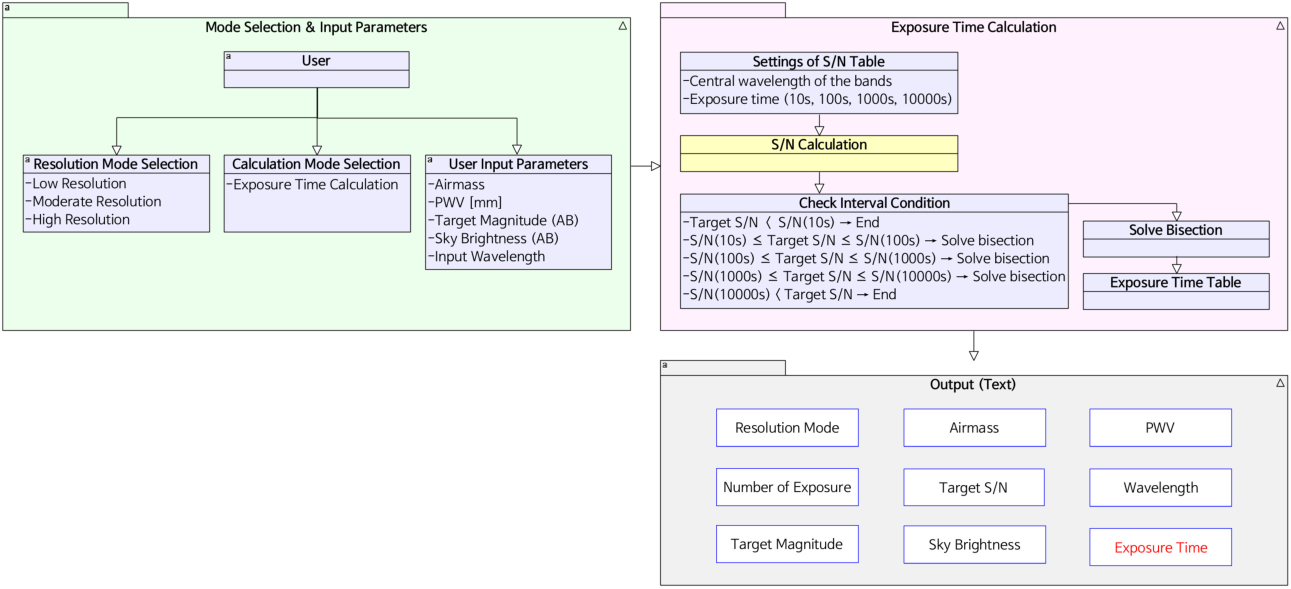}
   \caption[The flowchart shows the functions and parameters used in the exposure time calculation mode.]
   { \label{fig:fig_cal2} The flowchart shows the functions and parameters used in the exposure time calculation mode.}
   \end{figure}

For this calculation mode, we used the bisection method to estimate the exposure time for a single exposure, based on the S/N calculation. Figure~\ref{fig:fig_cal2} shows that the exposure time calculation employs the same parameters as the S/N mode, with the exception of certain user input parameters. Similar to the S/N calculation mode, the wavelength value is typically set to the central wavelength of the each band. However, there is an option for users to enter a specific wavelength if desired.

\begin{figure} [!ht]
\includegraphics[width=\textwidth]{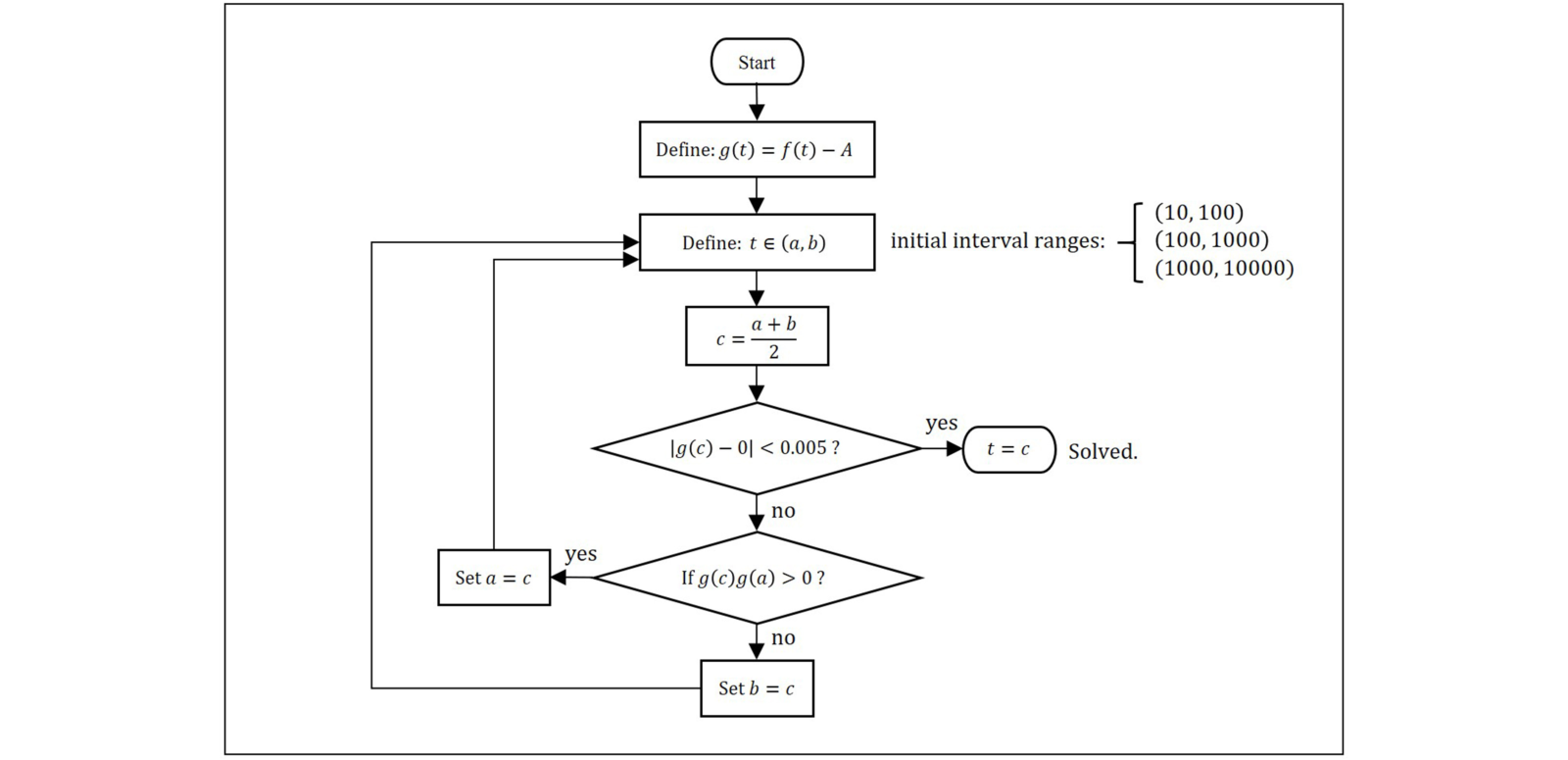}
\caption[Flowchart of the bisection method. $f(t)$ is the S/N as a function of the exposure time, $A$ is the target S/N. The initial interval ranges ensure the existence of at least one root of $g(t)$ in the interval. The convergence criterion is set to 0.005.]
{ \label{fig:bisection} Flowchart of the bisection method. $f(t)$ is the S/N as a function of the exposure time, $A$ is the target S/N. The initial interval ranges ensure the existence of at least one root of $g(t)$ in the interval. The convergence criterion is set to 0.005.}
\end{figure}

The bisection method is an iterative procedure that continuously reduces the search interval for a root by dividing the root interval into two equal subintervals, and then identifying which subinterval contains the root. Accordingly, the steps to derive the exposure time using this method are as follows Figure~\ref{fig:bisection}.

\subsection{S/N vs. Magnitude Calculation}
\begin{figure} [h]
   \includegraphics[width=\textwidth]{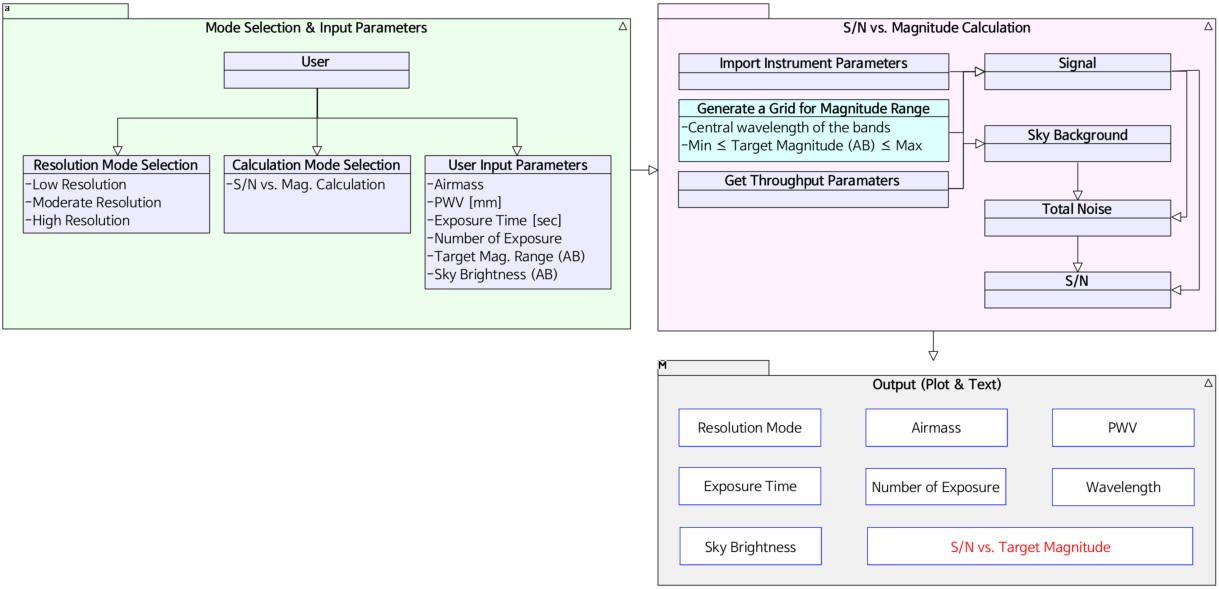}
   \caption[The flowchart of functions and parameters for S/N vs. Magnitude calculation mode.] 
   { \label{fig:fig_cal3} The flowchart of functions and parameters for S/N vs. Magnitude calculation mode.}
   \end{figure}

This mode produces a plot of S/N versus target magnitude in terms of the exposure time and the number of exposures. The first step in this process involves creating a grid array of target magnitudes within the user-defined magnitude range with steps of 0.1 mag. Subsequently, a grid of S/N values is generated through the step-by-step sequential input of these values into the S/N calculation. Figure~\ref{fig:fig_cal3} shows the detailed flowchart for this mode.

   \begin{figure} [b]
   \includegraphics[width=\textwidth]{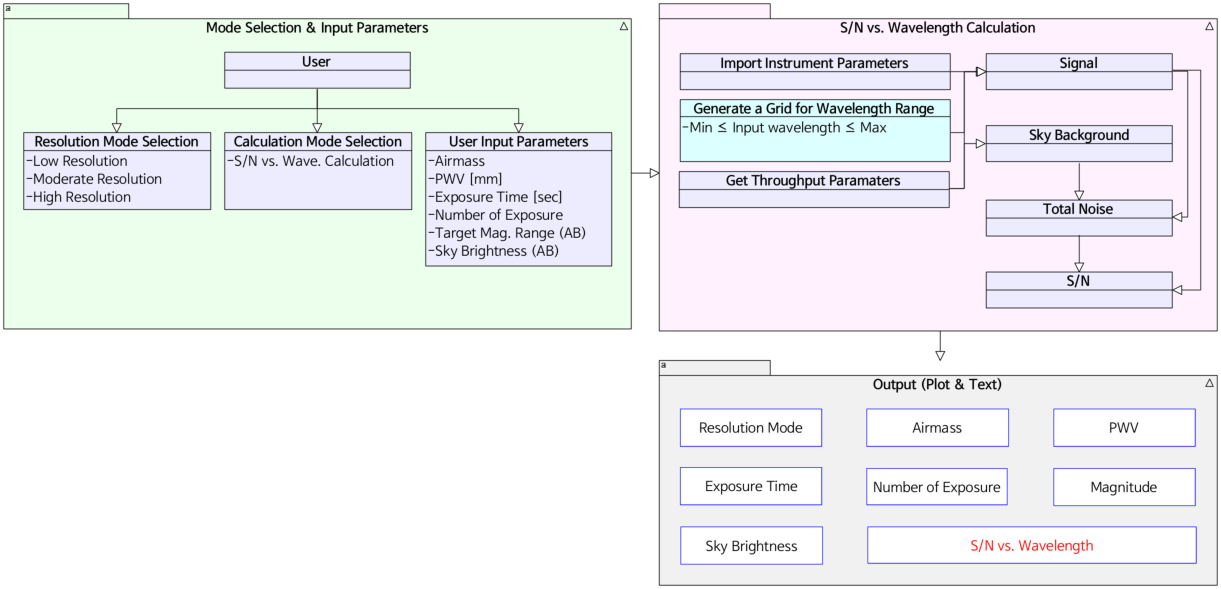}
   \caption[The flowchart of functions and parameters for S/N vs. Wavelength calculation mode.] 
   { \label{fig:fig_cal4} The flowchart of functions and parameters for S/N vs. Wavelength calculation mode.}
   \end{figure}

\subsection{S/N vs. Wavelength Calculation}
This mode produces a plot of S/N versus wavelength in terms of exposure time, the number of exposures, and the target magnitude. Initially, a grid array of wavelengths is created within the user-selected band and is comprised of 0.1 nm steps. Next, the wavelength grid of points is sequentially input into the S/N calculation (with the given parameters), generating a grid of S/N values. Figure~\ref{fig:fig_cal4} displays the detailed flowchart for this mode.

\section{Software}
\label{sec:software}

\subsection{Development process}
The software development process of the MSE ETC is based on the Agile development methodology. Agile code development is flexible in nature with software improvements implemented in each development cycle as a result of user feedback (as well as incorporation of any changing requirements).

We define the development cycle of the MSE ETC as an iteration from design to release for each calculation mode. During the design phase, the development document compiles MSE requirements, calculation methods, and software structure. We employ Visual Paradigm 17.1 (Visual Paradigm International Ltd.) as a modeling tool for Unified Modeling Language (UML) to make flowcharts and architectures. The developers proceed to implement the GUI and features using an integrated development environment (IDE) such as PyCharm or Microsoft Visual Studio Code. We manually perform white-box testing to trace the data flow within units of functions and classes, which also includes integration testing for both the GUI and multiple classes.

We use the Git repository for software version control. The developers create a local repository by forking from the origin (master) repository, and development takes place on the local computer using the cloned software from the local repository. After making changes to the software, the developers push the updated version to the local repository and submit a pull request to merge it into the origin repository. The project manager approves the merge after checking the code differences and reviewing its functionality. If there are changes in requirements, feedback from users, or bugs, the project manager notifies an issue using the GitHub Issue Tracker. Subsequently, the developer checks the issue and initiates the next development cycle.

\subsection{Software architecture}
   
\begin{figure} [b]
   \includegraphics[width=\textwidth]{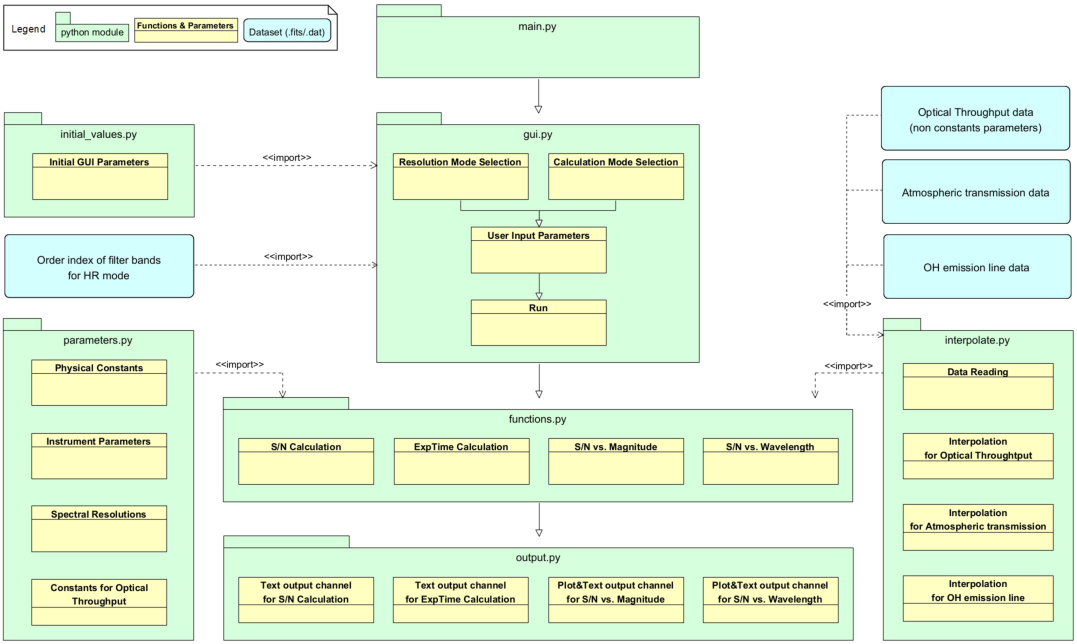}
   \caption[The software architecture of the MSE ETC code. The figure shows the connection among the various Python modules and the general flow of the code.  The dashed lines represents how data/derived values are imported from other modules.] 
   { \label{fig:fig_architecture} 
The software architecture of the MSE ETC code. The figure shows the connection among the various Python modules and the general flow of the code.  The dashed lines represents how data/derived values are imported from other modules.}
   \end{figure}

The MSE ETC code consists of six modules (with main.py serving as the main/overarching module; see Figure.~\ref{fig:fig_architecture}). The GUI and initial parameter settings are found in the gui.py and    
initial\_values.py routines, respectively. The MSE instrument parameters as well as select physical constants are defined in parameters.py. Throughput and atmospheric transmission determinations are performed by the interpolate.py routine. All input and derived parameters are then fed to functions.py, which performs the four main calculation modes. Finally, the results are displayed via output.py.

\subsection{Graphical user interface (GUI)}
The GUI of the MSE ETC is designed with the Tkinter library in Python for cross-platform use.
   \begin{figure} [b]
   \includegraphics[width=\textwidth]{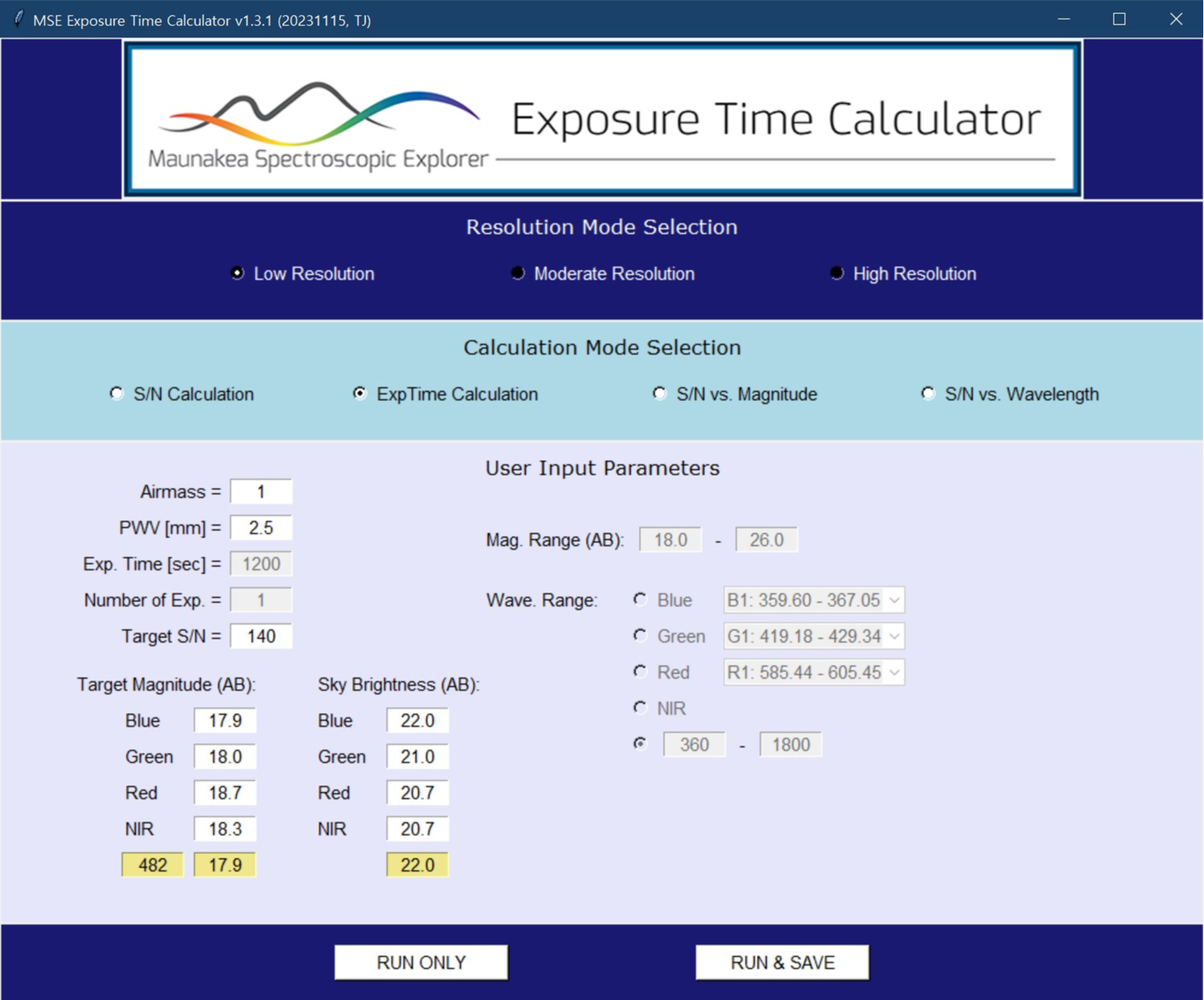}
   \caption[The GUI of the MSE ETC. The GUI is relies upon the Python 3 Tkinter library and a has pop-up style display.] 
   { \label{fig:fig_gui}
The GUI of the MSE ETC. The GUI is relies upon the Python 3 Tkinter library and a has pop-up style display.}
   \end{figure}
Accordingly, the ETC is able to run on Windows, Linux, and Mac OS systems. The instrument parameters are input automatically when the user selects the spectral resolution and calculation mode. The results of the calculation are displayed in new windows: a terminal shows the output results and (when selected) a graphical display that shows associated plots of the data (see Figure ~\ref{fig:fig_gui}).

\section{Test and Results}
\label{sec:results}

\begin{figure} [!ht]
   \includegraphics[width=\textwidth]{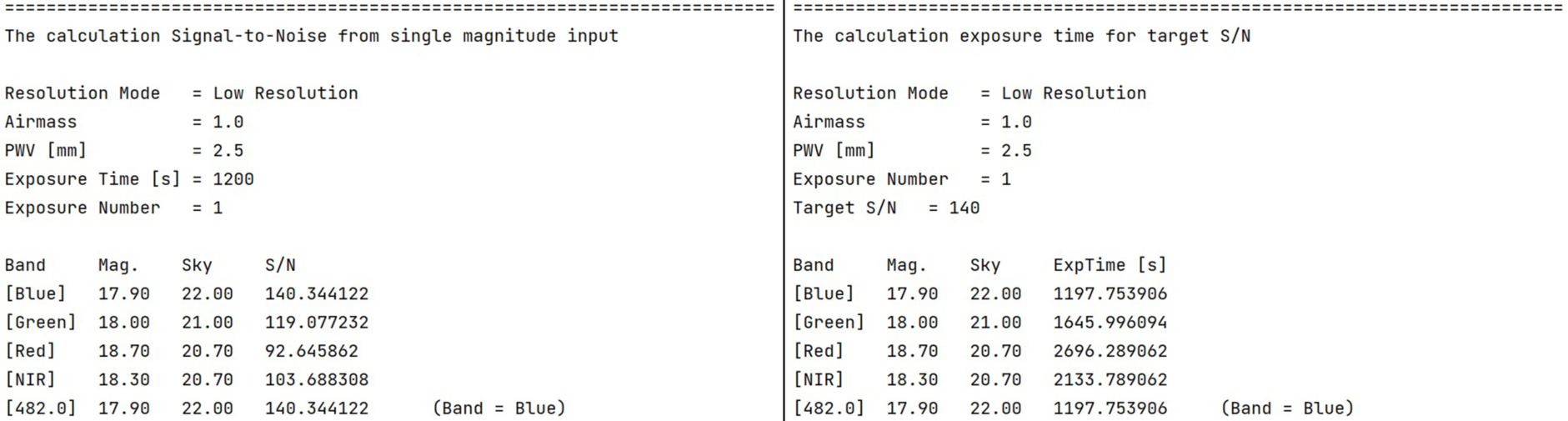}
   \caption[Results of the S/N calculation mode (left) and the exposure time calculation mode (right) for single target magnitude input in the LR spectrograph. Blue, Green, Red, and NIR indicate the central wavelength of the bands.] 
   { \label{fig:fig_result1} Results of the S/N calculation mode (left) and the exposure time calculation mode (right) for single target magnitude input in the LR spectrograph. Blue, Green, Red, and NIR indicate the central wavelength of the bands.}
   \end{figure}

The results of the S/N calculation and the exposure time calculation mode are displayed in Figure~\ref{fig:fig_result1}. In this test, we set LR as the resolution mode, airmass to 1.0, PWV to 2.5 mm, and the number of exposures to 1. The target magnitude and sky brightness values are input differently for testing, with a range of 17.90 to 18.70 mag for the target and 20.70 to 22.0 mag/$\text{arcsec}^{2}$ for the sky, respectively. (Note that it is expected to offer the capability to input a template spectrum for in a future update.) In an analogous fashion, we tested exposure time calculation mode under the same conditions with the signal to noise set to S/N=140 in the blue wavelength band for the LR instrument. The derived exposure time was 1,200 seconds, equivalent to the value found in the S/N calculation mode.

   \begin{figure} [!ht]
   \begin{center}
   \includegraphics[width=\textwidth]{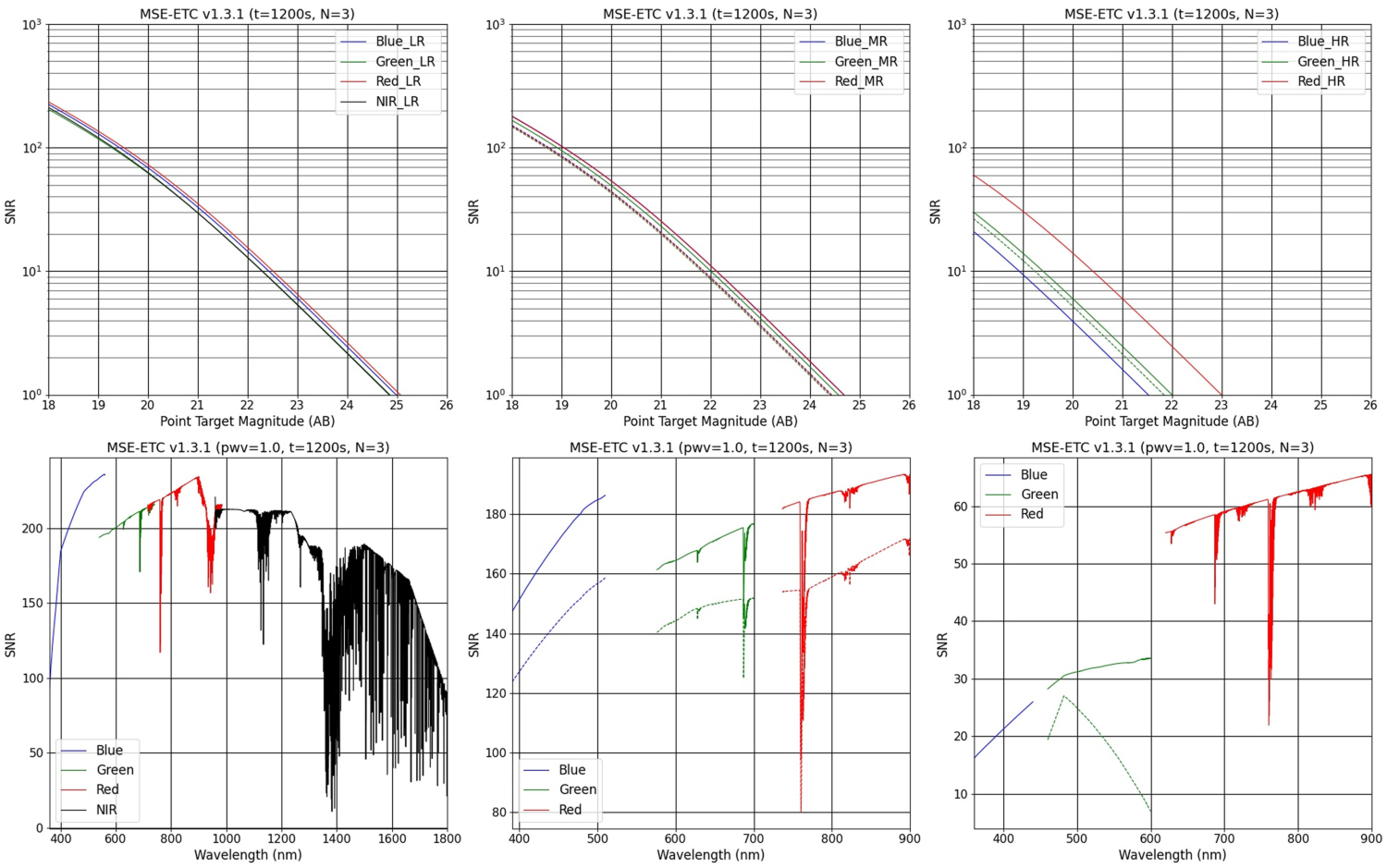}
   \caption[Plots of the output from the S/N vs. Magnitude calculation mode (top) and the S/N vs. Wavelength calculation mode (bottom). Data is generated for the LR (left), MR (center), and HR (right) spectral resolution settings. For the set of S/N vs. Magnitude and S/N vs. Wavelength plots, the dashed lines in the MR and HR modes show the results which also include the grating efficiency. Note that the grating efficiency of the HR spectrograph is currently available as a single value at 482 nm in the Green band.] 
   { \label{fig:fig_result2} Plots of the output from the S/N vs. Magnitude calculation mode (top) and the S/N vs. Wavelength calculation mode (bottom). Data is generated for the LR (left), MR (center), and HR (right) spectral resolution settings. For the set of S/N vs. Magnitude and S/N vs. Wavelength plots, the dashed lines in the MR and HR modes show the results which also include the grating efficiency. Note that the grating efficiency of the HR spectrograph is currently available as a single value at 482 nm in the Green band.}
   \end{center}
   \end{figure}

The results from the S/N vs. Magnitude and S/N vs. Wavelength calculation modes are displayed as plots as well as stored in data arrays.  Figure~\ref{fig:fig_result2} features sample outputs from these two calculation modes. As shown in this Figure, we have set the airmass to 1.0, PWV to 1.0 mm, exposure time to 1,200 seconds, and the number of exposures to 3. The target and sky brightness magntiudes were set to 18.00 mag and 20.70 mag/$\text{arcsec}^{2}$, respectively. Note that the telluric emission from OH lines is clearly detectable in NIR wavelength band (as shown in the S/N vs. Wavelength mode results for the LR setting).

In the signal calculation (see Equation (\ref{eq_signal})), where the target magnitude is constant, the signal size is determined by the ratio of total throughput ($\tau_{\text{total}}$) to spectral resolution ($R$). In LR mode, the $\tau_{\text{total}}/R$ for each band are Blue = 0.576, Green = 0.482, Red = 0.625, and NIR = 0.525, respectively. Therefore, the signal size differs by -16\% in the Green band, +8\% in the Red band, and -8\% in the NIR band compared to the Blue band. Similarly, in MR mode (without grating), the signal size differs by -13\% in the Green band and -1\% in the Red band compared to the Blue band. Notably, the Red band has only a 1\% difference from the Blue band, and the curves for the Blue and Red bands appear to entirely overlap in the S/N vs. magnitude plot in MR mode. Additionally, in HR mode (without grating), the signal differences are +60\% in the Green band and +340\% in the Red band compared to the Blue band. Consequently, the differences in S/N for each band of LR and MR modes seem to be small.

We estimated the processing time of MSE ETC for both exposure time calculation and S/N vs. Wavelength calculation in the LR mode. The testing was conducted using a laptop with an Intel Core i5-6200U CPU @ 2.30 GHz. As a result, the average processing time is ~7 seconds and ~111 seconds, respectively. In the future, as the MSE ETC handles multiple targets, improvements are necessary to reduce the processing time, taking into account its impact on the entire end-to-end survey planning.

\section{Conclusion}
\label{sec:conclusion}

We have developed the MSE ETC code to simulate the instrument performance of the MSE system, which will consist of an LMR spectrograph operating at two resolution settings (LR$\sim$3,000 and MR$\sim$6,000) and an HR spectrograph operating at one resolution (HR$\approx$30,000).  We employ the Agile methodology to ensure a flexible software development process. We also use a Git repository for code deployment and version tracking/management. In this paper, we have described the methodology and associated parameters for the four calculation modes of the MSE ETC. We have created a code architecture that allows for easy understanding of the data flow and code structure. We have also performed simulations and verified the results for each calculation mode of the ETC. In the future, we will implement additional code flexibility (e.g., consideration of extended source targets) as well as various software upgrades. Moreover, we will incorporate user feedback (e.g., from the MSE science user community). All of these code enhancements will be reflected in future releases. 

\subsection*{Code, Data, and Materials Availability}
The MSE ETC code is publicly available on GitHub (\href{https://github.com/mse-cfht/etc_khu_group}{https://github.com/mse-cfht/etc\_khu\_group}).

\subsection* {Acknowledgments}
This work was supported by the International Research \& Development Program of the National Research Foundation of Korea (NRF) funded by the Ministry of Science and ICT (Grant number: 2020K1A3A1A2104184711). S.E.H. and M.Y. were supported by the project ``Understanding Dark Universe Using Large Scale Structure of the Universe'', funded by the Ministry of Science and ICT, South Korea.

This manuscript contains the scientific content previously reported in SPIE proceeding at Astronomical Telescopes + Instrumentation 2022. Tae-Geun Ji, Taeeun Kim, Changgon Kim, Hojae Ahn, Mingyeong Yang, Soojong Pak, Sungwook E. Hong, Jennifer E. Sobeck, Kei Szeto, Jennifer L. Marshall, Christian Surace, "An exposure time calculator for the Maunakea Spectroscopic Explorer," Proc. SPIE 12189, Software and Cyberinfrastructure for Astronomy VII, 121892N (29 August 2022); https://doi.org/10.1117/12.2629101 \cite{Ji2022}

\bibliography{spiebib}   
\bibliographystyle{spiejour}   


\vspace{2ex}\noindent\textbf{Tae-Geun Ji} is a graduate student at the Kyung Hee Univercity (KHU), Republic of Korea. He received has BS degrees in astronomy and space science from the KHU in 2016. His current research interests include astronomical instrumentation, software engineering for ground-based optical telescopes.

\vspace{2ex}\noindent\textbf{Sungwook E. Hong} is a senior researcher at the Korea Astronomy and Space Science Institute (KASI), Republic of Korea. He is also a associate professor at the Astronomy Campus of the University of Science and Technology (UST). His current research interests include large-scale structure of the universe, cosmological simulations, deep learning in astronomy, astrobiology, cosmic reionization, and astronomical instrumentation.

\vspace{2ex}\noindent\textbf{Changgon Kim} is a graduate student at Kyung Hee University, South Korea. He received his BS degree in astronomy and space science from the same university in 2021. His current research interests include astronomical instrumentation, control software, infrared optical design, optical photon simulation, and optical telescopes for space satellites.

\vspace{1ex}
\noindent Biographies and photographs of the other authors are not available.

\listoffigures
\listoftables

\end{spacing}
\end{document}